\documentclass[12pt, a4paper]{article}
\usepackage{tikz}
\usepackage[dvips]{epsfig}
\usepackage{graphics}
\usepackage{verbatim}
\usepackage{latexsym}
\usepackage{amssymb} 
\usepackage{mathrsfs}

\usepackage[vlined,titlenumbered]{algorithm2e}

\newtheorem{Definition}{Definition}[section]
\newtheorem{Theorem}[Definition]{Theorem}
\newtheorem{Lemma}[Definition]{Lemma}

\newtheorem{Corollary}[Definition]{Corollary}

\newtheorem{Property}[Definition]{Property}
\newtheorem{Proposition}[Definition]{Proposition}

\newtheorem{Characterization}[Definition]{Characterization}
\newtheorem{Decomposition Step}[Definition]{Decomposition Step}
\newtheorem{Notation}[Definition]{Notation}
\newenvironment{Proof}{\noindent {\bf Proof:} }{{ {$\Box$}}\vspace{\baselineskip}}

\newtheorem{Example}[Definition]{Example}

\newcommand{\cms}{clique minimal separator~}

\newcommand{\cmss}{clique minimal separators~}

\newcommand{\ms}{minimal separator~}

\newcommand{\mt}{minimal triangulation~}

\newcommand{\cmsd}{clique minimal separator decomposition~}

\newcommand{\mtn}{minimal triangulation}

\title{Computing the atom graph of a graph and the union join graph of a hypergraph}
\author{Anne Berry\thanks{LIMOS UMR CNRS 6158, Ensemble Scientifique
des C\'ezeaux, F-63 173 Aubi\`ere, France, berry@isima.fr}
\addtocounter{footnote}{2}
\and Genevi\`eve Simonet\thanks{LIRMM, 161 Rue Ada, F-34392 Montpellier,
France, genevieve.simonet@univ-montp2.fr}
}
\begin{document}

\tikzset{ens/.style={draw, rectangle,rounded corners=5pt}, 
              dot/.style={dotted} } 

\maketitle
%%%%%%%%%%%%%%%%%%%%%%%%%%%%%%%%%%%%%%%%%%%%%%%%%%%%%%%%%%%%%%%%%%%%%%%%
\begin{abstract}
The atom graph of a graph is the graph whose vertices are the atoms obtained by clique minimal separator decomposition of this graph, and whose edges are the edges of all possible atom trees of this graph. We provide two efficient algorithms for computing this atom graph, with a complexity in $O(min(n^\alpha \log n, nm, n(n+\overline{m}))$ time, 
which is no more than the complexity of computing the atoms in the general case.
%\par
We extend our results to $\alpha$-acyclic hypergraphs.  We introduce the notion of union join graph, which is the union of all possible join trees; we apply our algorithms for atom graphs to efficiently compute union join graphs. \\

\textbf{Keywords: } clique separator decomposition, atom tree, atom graph, clique tree, clique graph, $\alpha$-acyclic hypergraph.
\end{abstract}

%%%%%%%%%%%%%%%%%%%%%%%%%%%%%%%%%%%%%%%%%%
\section{Introduction}\label{introduction}
%%%%%%%%%%%%%%%%%%%%%%%%%%%%%%%%%%%%%%%%%%
Decomposition by 
\cmss (into subgraphs called \textit{atoms}) was
introduced by Tarjan \cite{Tarjan85}
in 1985 
as a useful hole- and antihole-preserving decomposition. It turns out that this decomposition is unique when clique \textit{minimal}
separators are used \cite{Leimer}.
\par
This decomposition has given rise to recent interest, both in the general case \cite{introduction_cms,AtomTree,CoudertDucoffHAL,Leimer}
and for special graph classes \cite{HDfree,Clawfreedecomp,paraglider,BraHoa2005,BraLeMah2007,Cliqueorhole}.
Applications have arisen in the fields of databases \cite{Olesen}, 
text mining \cite{kaba_didi} and biology \cite{Mosaic,InSilico}. 
\par
Recently, \cite{AtomTree} introduced the concept of \textit{atom tree}, which organizes the atoms of the \cmsd into a tree as a generalization of 
the clique tree for chordal graphs: the nodes are the atoms, and the edges correspond to the \cmss of the graph. 
However, as is the case for the clique tree, the atom tree is not uniquely defined. This can be a problem for instance
with the promising use of an atom tree as a visualization tool.
\par
In this paper, we focus on the \textit{atom graph}, whose vertices are the atoms, and whose edges are those of all possible atom trees.
\par
Atom graphs have been used in several papers.
\par
In the general case, the notion of atom graph was introduced in 2007 in \cite{InSilico} in the context of visualizing biological clusters.  
An efficient construction algorithm was proposed in 2010 in \cite{HAL2010}.
\par
In the case of chordal graphs, the atoms are the maximal cliques and the atom trees are the clique trees. A related graph which has been extensively studied in this context 
is the \textit{clique graph}  (see e.g. \cite{Celina_clique_graph,CliqueGraphRecPol}), which is the intersection graph of the maximal cliques.
The \textit{weighted clique graph} of a chordal graph has been used to construct a clique tree \cite{BP,Gavril}: any maximum weight spanning tree is a clique tree, and vice-versa.
Thus, except for some very special cases, the atom graph is a proper subgraph of the clique graph.
% , as illustrated by 
% Example \ref{Ex:cliquegraph}.
In the context of efficiently constructing a clique tree, in 1991 \cite{BP} studied the family of all possible clique trees, an object very close to the atom graph of a chordal graph. 
In 1995, \cite{Habib1} used the weighted atom graph of a chordal graph, but misguidedly called this object the 'clique graph'. 
In 2012 in \cite{Habib2}, this object is further studied and called the 'reduced clique graph'.
\par
Our first goal in this paper is to propose efficient algorithms to compute the atom graph, both in the general case and in the case of chordal graphs.
\par
Given a graph, all known algorithms for computing the decomposition into atoms first compute a \mt of the graph \cite{introduction_cms,AtomTree,Leimer}, 
with the exception of some special graph classes \cite{HDfree,Clawfreedecomp}. A \mt can be computed in $O(min(n^{\alpha} \log n ,nm,n(n+\overline{m})))$ time 
where $\alpha$ is the real number such that $O(n^{\alpha})$ is the best known time complexity for matrix multiplication
and $\overline{m}$ is the number of edges of the complement of $G$
\cite{AtomTree,Pinar_TM,RTL}. 
From this \mtn, an atom tree can be computed in $O(min(n^{\alpha}, nm, n(n+t)))$ 
time \cite{AtomTree,CoudertDucoffHAL,Leimer}, where $t$ is the number of 2-pairs of the \mtn, and thus $t \leq \overline{m}$. 
As a result, an atom tree can be computed in $O(min(n^{\alpha} \log n, nm,n(n+\overline{m})))$
time.
\par
To compute the atom graph efficiently, we present two different approaches.
One takes as input an atom tree as well as the inclusion relation between the separators represented by its edges, 
and the other takes as input the weighted intersection graph of the atoms. In both cases, we provide an $O(n^2)$ algorithm to compute the atom graph from the input.  
Our global complexity when taking the graph itself as input comes to $O(min(n^{\alpha} \log n, nm,n(n+\overline{m})))$ time.
\par
We then go on to remark that the atoms of a graph $G=(V,E)$ 
can be seen as the hyperedges of an $\alpha$-acyclic hypergraph, whose vertex set is $V$, 
since $G$ has an atom tree that is a join tree of this hypergraph.
However, the atoms of a graph are pairwise non-inclusive, which is not a requirement for $\alpha$-acyclic hypergraphs, where a hyperedge can be 
included in another. Fortunately, our algorithms also work in this more general context.
\par
We introduce the notion of \textit{union join graph}, which is the union of all join trees, and provide algorithms to compute this object 
efficiently.
\par
The paper is organized as follows:
Section 2 provides some necessary preliminaries.
Section 3 discusses useful properties of the atom graph.
Section 4 presents our algorithms to compute the atom graph.
Section 5 defines the atom hypergraph and relates it to $\alpha$-acyclic hypergraphs.
Section 6 discusses how to compute the union join graph of an $\alpha$-acyclic hypergraph.
We conclude in Section 7.

%%%%%%%%%%%%%%%%%%%%%%%%%%%%%%%%
\section{Preliminaries}\label{2}
%%%%%%%%%%%%%%%%%%%%%%%%%%%%%%%%
The graphs considered in this paper are finite and undirected.
For a graph $G=(V,E)$, $n=|V|$ and $m=|E|$. 
For any subset $X$ of $V$, $G(X)$ denotes the subgraph of $G$ induced by $S$.
For any vertex $v$ of $G$, $N_G(v)$ denotes the neighborhood of $v$ in $G$: $N_G(v) = \{w \in V~|~vw \in E\}$.
We will omit the subscripts when there is no ambiguity. 
A {\it clique} of $G$ is a set of pairwise adjacent
vertices of $G$, and $G$ is {\it complete} if $V$ is a clique of $G$.
%
%We will say that we {\it saturate} a set $S$ of vertices when we add to the 
%graph all the edges necessary to make $S$ into a clique.
% 
The \textit{union of two graphs} $G_1=(V,E_1)$ and $G_2=(V,E_2)$ is $G_1 \cup G_2 = (V, E_1 \cup E_2)$.
\par
$\overline{G}$ denotes the \textit{complement} of $G$, and $\overline{m}$ denotes its number of edges. 
$\alpha$ is the real number such that $O(n^{\alpha})$ is the best known time complexity for matrix multiplication. 
For any set $V$, ${\cal{P}}(V)$ is the power set of $V$.
For any subset ${\cal A}$ of ${\cal{P}}(V)$, the \textit{intersection graph} of ${\cal A}$ 
is the graph $({\cal A},E)$ where $E$ is the set of pairs of ${\cal A}$ whose intersection is non-empty.
For each graph $G$, ${\cal K}(G)$ denotes the set of maximal cliques of $G$ and the \textit{clique graph} of $G$, 
is the intersection graph of ${\cal K}(G)$.
If $X$ and $Y$ are nodes of a tree $T$, $P_T(X,Y)$ denotes the path in $T$ between $X$ and $Y$. 
%

%%%%%%%%%%%%%%%%%%%%%%%%%%%%%%%%
\par
\noindent
%§§§§§§§§§§§§§§§§§§§§§§§§§§§
\textbf{\textit{Separation.}}
%§§§§§§§§§§§§§§§§§§§§§§§§§§§
Let $S$ be a subset of vertices of a connected graph $G=(V,E)$. 
$S$ is a {\em separator} of $G$ if $G(V \setminus S)$ is disconnected.
For any vertices $a$ and $b$ in $V \setminus S$,
$S$ is an {\em $ab$-separator} of $G$
%
%, or $S$ {\em separates} $a$ and $b$ in $G$, 
%
if $a$ and $b$ are in different connected components of $G(V \setminus S)$. 
$S$ is a {\em minimal $ab$-separator} if it is an inclusion-minimal $ab$-separator,
and a \emph{minimal separator} if there is some pair $\{ a,b\}$ of vertices such that $S$ is a
minimal $ab$-separator. 
Given a \ms $S$, $C$ is a \textit{full component} of $S$ if $C$ is a connected component of $G(V \setminus S)$ and $N_G(C)=S$.
$S$ is a \ms if $S$ has at least 2 full components, and $S$ is a minimal ab-separator if $a$ and $b$ lie in 2 different full components of $S$.
Given three subsets $S$, $A$ and $B$ of $V$, $S$ is a (minimal) $AB$-separator of $G$ if it is a (minimal) $ab$-separator of $G$ for each $a \in A$ and each $b \in B$.
\par
A \textit{2-pair}  of a connected graph $G$ is a pair $\{ x,y \}$
of non-adjacent vertices such that every chordless path between $x$ and $y$ is of length 2, or 
equivalently such that $N(x) \cap N(y)$ is a minimal $xy$-separator of $G$.  The number of 2-pairs of a graph is denoted $t$, 
with $t \leq \overline{m}$.

\par
If $G$ is disconnected then a (minimal) ($ab$-)separator of $G$ is a (minimal) ($ab$-)separator of one of its connected components. Thus the set of minimal separators of a graph is the union of the sets of minimal separators of its connected components, and so it is for its set of 2-pairs.
%
% Given two graphs $G=(V,E)$ and $G'=(V,E')$, and a subset $S$ of $V$, if 
% $S$ has the same components in $G$ as in $G'$, and if, for each such component $C$, 
% $N_G(C)=N_{G'}(C)$, 
% we say that $S$ has \emph{the same components and neighborhoods} in $G$ and $G'$.
%
%%%%%%%%%%%%%%%%%%%%%%%%%%%%%%%
\par
\noindent
%§§§§§§§§§§§§§§§§§§§§§§§§§§§
\textbf{\textit{Chordal graph}s.}  
%§§§§§§§§§§§§§§§§§§§§§§§§§§§
A graph is {\it chordal}, or {\it triangulated}, if it has no chordless 
cycle of length at least $4$. 
A graph is chordal if and only if all its minimal separators are cliques \cite{Dirac}.
A chordal graph has at most $n$ maximal cliques and the sum of their sizes is bounded by $n+m$.
A connected graph is chordal if and only if it has a \textit{clique tree} \cite{Buneman,Gavril}.

\begin{Definition}
Let $G=(V,E)$ be a connected chordal graph. 
A {\em clique tree} of $G$ is a tree $T = ({\cal K}(G),E_T)$ such that
for each vertex $x$ of $G$, the set ${\cal K}_x$ of nodes of $T$ containing $x$ induces a subtree of $T$.
\end{Definition}

\begin{Characterization}\label{carEdgeCliqueTree}
\cite{BP}
Let $G=(V,E)$ be a connected chordal graph, let $T$ be a clique tree of $G$, and let $S \subseteq V$; then $S$ is a minimal separator of $G$ if and only if 
there is an edge $XY$ of $T$ 
such that $S=X \cap Y$.
\end{Characterization}

If $G$ is a disconnected chordal graph, we associate with $G$ a forest whose connected components are clique 
trees of the connected components of $G$. 
A clique tree (forest) can be computed in linear time \cite{BP}. 
\par
\noindent
\textbf{\textit{Atoms.}}
Atoms are the subgraphs obtained by applying the decomposition by \cmss (see \cite{introduction_cms} for a survey). 

\begin{Characterization}\cite{Leimer}
An atom of a graph $G = (V,E)$ is an inclusion-maximal subset of $V$ inducing a connected subgraph of $G$ with no 
clique separator. 
\end{Characterization}

We will denote the set of atoms of $G$ by ${\cal A}(G)$.

\begin{Property} \label{cgatoms}
 The atoms of a chordal graph are its maximal cliques.
\end{Property}

\begin{Property} \label{propInterAtoms} \cite{Leimer}
The intersection of two distinct atoms is a clique.
\end{Property}

\begin{Notation} \label{notGplus} 
For a graph $G = (V,E)$, $G^+$ denotes the graph whose vertex set is $V$ and whose edges are the pairs of $V$ that 
are contained in a common atom of $G$ (this graph is denoted $G^*$ in \cite{Leimer}).
\end{Notation}

\begin{Property} \label{propG*Leimer} \cite{Leimer}
For a graph $G$, $G^+$ is chordal, its maximal cliques are the atoms of $G$ and for each 
clique $S$ of $G$ and each pair $\{a,b\}$ of $V \setminus S$, $S$ is an $ab$-separator (resp. minimal $ab$-separator) of $G$ if and only if  $S$ is an $ab$-separator (resp. minimal $ab$-separator) of $G^+$.
\end{Property}

It follows that a graph has at most $n$ atoms.

\par
\noindent
%§§§§§§§§§§§§§§§§§§§§§§§§§§§
\textbf{\textit{\textit{Atom trees}.}}
%§§§§§§§§§§§§§§§§§§§§§§§§§§§

To represent the atoms of a graph, \cite{AtomTree} extend the notion of clique tree of a connected chordal graph to 
the notion of \textit{atom tree} of a connected graph:

\begin{Definition}\cite{AtomTree}
Let $G=(V,E)$ be a connected graph. 
An {\em atom tree} of $G$ is a tree $T = ({\cal A}(G),E_T)$
such that for each vertex $x$ of $G$, the set ${\cal A}_x$ of nodes of $T$ containing $x$ induces a subtree of $T$.
\end{Definition}

Note that an atom tree is not a decomposition tree of clique separator decomposition as defined in \cite{HAL2010,Tarjan85}, thouh this deomposition is called `atom tree' in \cite{HAL2010}.
\par
An atom tree of a connected graph $G$ can be computed in $O(min(n^{\alpha} \log n, nm,n(n+\overline{m})))$ time \cite{AtomTree,CoudertDucoffHAL}.

The edges of an atom tree of a graph correspond to its clique minimal separators.
\begin{Characterization}\label{carEdgeAtomTree}
\cite{AtomTree}
Let $G=(V,E)$ be a connected graph, let $T$ be an atom tree of $G$, and let $S \subseteq V$; then $S$ is a clique minimal separator of $G$ if and only if 
there is an edge $AB$ of $T$ 
such that $S=A \cap B$.
\end{Characterization}

%\begin{Property}\label{Prop:numberoccurS}
 %Let $G=(V,E)$ be a connected graph, let $T$ be an atom tree of $G$, and let $S$ be a \cms of $G$.  Then 
 %the number of edges of $T$ which correspond to $S$ is equal to the number of full components of $S$ minus 1.
%\end{Property}

\begin{Property} \label{propG*}
 For a connected graph $G$, the atom trees of $G$ are the clique trees of the chordal graph $G^+$ 
%obtained by saturating the atoms of $G$. 
($G^+$ is defined in Notation~\ref{notGplus}).
\end{Property}

\begin{figure}
\begin{center}
\begin{tikzpicture} 

\begin{scope}
\coordinate (1) at (2,0) ;
\draw (1) node {$\bullet$}
               node  [above left] {$1$};

\coordinate (2) at (3,1) ;
\draw  (2) node {$\bullet$}
               node [left] {$2$};
               
 \coordinate (3) at (2,2) ;
\draw  (3) node {$\bullet$}
               node  [below left] {$3$};
               
\coordinate (4) at (3,3) ;
\draw  (4) node {$\bullet$}
               node  [right] {$4$};
                
\coordinate (5) at (0,2) ;
\draw  (5) node {$\bullet$}
               node  [left] {$5$};          
                
\coordinate (6) at (0,0) ;
\draw  (6) node {$\bullet$}
               node  [left] {$6$};  
                           
\coordinate (7) at (4,1) ;
\draw  (7) node {$\bullet$}
               node [right]  {$7$};   
                           
\coordinate (8) at (4,0) ;
\draw  (8) node {$\bullet$}
               node  [right] {$8$}; 
                             
\coordinate (9) at (-0,-1.5) ;
\draw  (9) node {$\bullet$}
               node  [left] {$9$};
              
\coordinate (10) at (3,-1) ;
\draw  (10) node {$\bullet$}
               node  [above] {$10$};

\coordinate (11) at (3,-2) ;
\draw  (11) node {$\bullet$}
               node  [below] {$11$};

\coordinate (12) at (4,-2) ;
\draw  (12) node {$\bullet$}
               node  [below] {$12$};

\coordinate (13) at (4,-1) ;
\draw  (13) node {$\bullet$}
               node  [above] {$13$};

\draw (1) -- (2) -- (3) -- (4) -- (5) -- (6) -- (1) -- (7) -- (8) -- (1) -- (9) ;
\draw (4) -- (2) -- (7) -- (1) -- (3) -- (7) ;
\draw (4) -- (3) -- (5) ; 
\draw (10) -- (11) -- (12) -- (13) -- (10) ;
\draw (10) -- (1) -- (11) ;

\draw  (0,2.7) node {$A =\{1,2,3,4,5,6\}$};
\draw  (4.5,1.5) node {$B =\{1,2,3,7\}$};
\draw  (3.7,0.4) node {$C$};
\draw  (0.7,-0.7) node {$D$};
\draw  (2.8,-1.2) node {$E$};
\draw  (3.5,-1.6) node {$F$};
\end{scope}      
              
\begin{scope}[scale = 0.6,yshift = -247]
\node [ens] (A)  at (0,2) {$A$};
\node [ens] (B)  at (2,3) {$B$};
\node [ens] (C)  at (4,2) {$C$};
\node [ens] (D)  at (0,0) {$D$};
\node [ens] (E)  at (2,-1) {$E$};
\node [ens] (F)  at (4,0) {$F$};

\draw (A) -- (B)  node [midway,above left] {$\{1,2,3\}$}
                -- (C)  node [midway,above right] {$\{1,7\}$};
\draw (A) -- (D)  node [midway,left] {$\{1\}$}
                -- (E)  node [midway,below left] {$\{1\}$}
                -- (F)  node [midway,below right] {$\{10,11\}$};
\end{scope}

\begin{scope}[scale = 0.6,xshift = 220,yshift = -247]
\node [ens] (A)  at (0,2) {$A$};
\node [ens] (B)  at (2,3) {$B$};
\node [ens] (C)  at (4,2) {$C$};
\node [ens] (D)  at (0,0) {$D$};
\node [ens] (E)  at (2,-1) {$E$};
\node [ens] (F)  at (4,0) {$F$};

\draw (A) -- (B)  node [midway,above left] {$\{1,2,3\}$}
                -- (C)  node [midway,above right] {$\{1,7\}$};
\draw (D) -- (B)  node [midway,right] {$\{1\}$}
                -- (E)  node [midway,right] {$\{1\}$}
                -- (F)  node [midway,below right] {$\{10,11\}$};
%\draw  (0.6,0.3) node {$\{1\}$};
\end{scope}

\end{tikzpicture}
\end{center} 

\caption{A graph $G$ and two atom trees of $G$.}  \label{figAtomTree}

\end{figure}
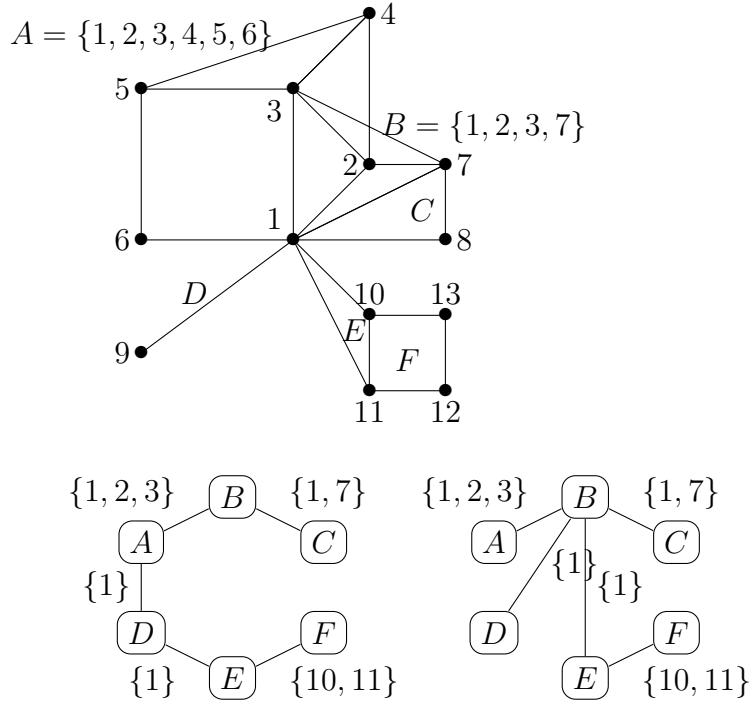

\begin{Example}
Figure~\ref{figAtomTree} shows a graph $G$ and two of its atom trees. 
The atoms of $G$ are $A = \{1,2,3,4,5,6\}$, $B = \{1,2,3,7\}$, $C = \{1,7,8\}$, $D = \{1,9\}$, $E = \{1,10,11\}$ and $F = \{10,11,12,13\}$.
Each edge $XY$ of each atom tree is labeled with the associated \cms $X \cap Y$ of $G$. 
$G$ has $15$ atom trees, which are all the trees obtained from the forest $({\cal A}(G), \{AB,BC,EF\})$ by adding $2$ edges not containing the node $F$.
\end{Example}

%\begin{figure}[h!]%\label{figAtomTree}
%\begin{center}
%\includegraphics[width=0.6\textwidth]{fig1}
%\end{center}
%\caption{A graph $G$ and two atom trees of $G$.}\label{figAtomTree}
%\end{figure}

%%%%%%%%%%%%%%%%%

%%%%%%%%%%%%%%%

% As for clique trees of chordal graphs, we can extend the definition of an atom tree to the definition 
% of an "atom forest" $F$ of an arbitrary graph $G$, the connected components of $F$ being atom trees 
% of the connected components of $G$, for which Characterization~\ref{carEdgeAtomTree} holds.

\noindent
The following properties will be used to compute complexity bounds. 

\begin{Property} \label{propDefAG}
Let $G$ be a connected graph and let $A$ and $B$ be distinct atoms of $G$. Then $G(A \setminus B)$ is connected and $A \cap B \subseteq N_G(A \setminus B)$.
\end{Property}

\begin{Proof} 
By Property~\ref{propInterAtoms} $A \cap B$ is a clique of $G$.
$A \setminus B$ is connected since otherwise $A \cap B$ would be a clique separator of $G(A)$.
Similarly $A \cap B \subseteq N_G(A \setminus B)$ since otherwise $A \cap N_G(A \setminus B)$ would be a clique $ab$-separator of $G(A)$ for any $a$ in $A \setminus B$ (which is non-empty by definition of atoms) and any $b$ in $(A \cap B) \setminus N_G(A \setminus B)$. 
\end{Proof}

\begin{Property} \label{propSomAtoms}
The sum of the sizes of the atoms of a graph is bounded by $n+m$.
\end{Property}

\begin{Proof}
It is sufficient to prove it in the case of a connected graph $G$.
Let $T$ be an atom tree of  $G$, and let us show by induction on $|{\cal A}|$ that for each connected subset ${\cal A}$ of nodes of $T$, $\Sigma_{X \in {\cal A}} |X| \leq n_{\cal A} + m_{\cal A}$, where $n{\cal A}$ and $m{\cal A}$ are the numbers of vertices and edges of the subgraph of $G$ induced by $V_{\cal A} = \cup_{X \in {\cal A}} X$.
 It trivially holds if $|{\cal A}|=1$.
We assume that it holds if $|{\cal A}| = k$. Let us show that it holds if $|{\cal A}| = k+1$. 
Let $X_1$ be a leaf of $T({\cal A})$, let $X_2$ be the neighbor of $X_1$ in $T({\cal A})$, let $S=X_1 \cap X_2$ and 
let ${\cal A}_2 = {\cal A} \setminus \{X_1\}$.
By induction hypothesis $\Sigma_{X \in {\cal A}_2} |X| \leq n_{{\cal A}_2} + m_{{\cal A}_2}$.
As $T$ is an atom tree of $G$, $V_{\cal A}$ is the disjoint union of $X_1 \setminus S$ and $V_{{\cal A}_2}$, so $n_{\cal A} = |X_1 \setminus  S| + n_{{\cal A}_2}$.
%
%As $S$ is a clique of $G$ by
%Property~\ref{propInterAtoms},
%$S \subseteq N_G(X_1 \setminus S)$ (otherwise $X_1 \cap N_G(X_1 \setminus S)$ would be a clique $xy$-separator of $G(X_1)$ for any $x$ in $X_1 \setminus S$ (which is non-empty by definition of atoms) and any $y$ in $S \setminus N_G(X_1 \setminus S)$, and $X_1$ would not be an atom of $G$). 
%Hence $m' \geq m_2 + |S|$.
%
By Property~\ref{propDefAG} $S \subseteq N_G(X_1 \setminus S)$, so $m_{\cal A} \geq |S| + m_{{\cal A}_2}$..
It follows that $\Sigma_{X \in {\cal A}} |X| = |X_1| + \Sigma_{X \in {\cal A}_2 } |X| \leq (|X_1 \setminus S| + |S|) + (n_{{\cal A}_2} + m_{{\cal A}_2}) = (|X_1 \setminus S|+n_{{\cal A}_2}) + (|S|+m_{{\cal A}_2}) \leq n_{\cal A} + m_{\cal A}$.
\end{Proof}
\begin{Property} \label{propSomMinSep}
The sum of the sizes of the sets $X \cap Y$ for each edge $XY$ of an atom tree $T$ of a graph is bounded by $n+m$, 
and these sets can be computed from $T$ in $O(m)$ time.
\end{Property}
\begin{Proof}
Let  $T = ({\cal A}(G), E_T)$ be an atom tree of $G$.
We consider a rooted directed tree $T_r = ({\cal A}(G), U)$  obtained from $T$ by choosing an arbitrary root.
Thus $\Sigma_{XY \in E_T} |X \cap  Y| = \Sigma_{(X,Y) \in U} |X \cap Y| \leq \Sigma_{(X,Y) \in U} |Y| \leq \Sigma_{Y \in {\cal A}(G)} |Y| \leq n+m$ by Property~\ref{propSomAtoms}. \\
These sets can be computed by searching $T$ and computing $X \cap Y$ in $O(|Y|)$ time when reaching $Y$ from its neighbor $X$, and therefore in $O(m)$ time.by Property~\ref{propSomAtoms}.
\end{Proof}

%\newpage
\par\noindent
%§§§§§§§§§§§§§§§§§§§§§§§§§§§
\textbf{$\alpha$-acyclic hypergraphs.}
%§§§§§§§§§§§§§§§§§§§§§§§§§§§
A \textit{simple hypergraph}, or \textit{hypergraph} for short, is a structure $H = (V,{\cal E})$, 
where $V$ is its vertex set and  ${\cal E}$ is a set of non-empty subsets of $V$, called the \textit{hyperedges} of $H$, whose union is equal to $V$.
% This last condition is not always required in the definition of a hypergraph given in the literature, 
% but it simplifies the formulation of some results and it is easy to extend them to hypergraphs not satisfying this condition.
A hypergraph is a \textit{clutter} if the elements of ${\cal E}$ are pairwise non-inclusive.
Its \textit{line graph}, denoted by $L(H)$, is the intersection graph of ${\cal E}$.
Its \textit{2-section graph}, denoted by $2SEC(H)$, is the graph whose vertex set is $V$ and whose edges are the pairs of $V$ that are contained in a hyperedge of $H$.
$H$ is \textit{connected} if $L(H)$ is connected, or equivalently if $2SEC(H)$ is connected. 
We denote by $p$ the number of hyperedges of a hypergraph.
% to distinguish it from the number $m$ of edges of a graph.. 
Let $(v_1, \ldots, v_n)$ be an ordering of $V$ and let  $(X_1, \ldots, X_p)$ be an ordering of ${\cal E}$. 
The \textit{incidence matrix} of $H$ w.r.t. these orderings is 
the $n \times p$ matrix $M = (m_{i,j})$ such that for each $i \in [1,n]$ and each $j \in [1,p]$, $m_{i,j} = 1$ if $v_i \in X_j$ and $0$ otherwise.
\par
A \textit{join tree} of $H$ is a tree $T$ whose node set is  ${\cal E}$ and such that for each vertex $x$ of $H$, 
the set ${\cal E}_x$ of nodes of $T$ containing $x$ induces a subtree of $T$, or equivalently, such that for each pair $\{X,Y\}$ of ${\cal E}$, $X \cap Y$ is a subset of each node of $P_T(X,Y)$.
$H$ is \textit{$\alpha$-acyclic} if it has a join tree.
%
%We will call \textit{$\alpha$-structure} a structure $(H,T)$ where $H$ is an $\alpha$-acyclic hypergraph and $T$ is a join tree of $H$. An $\alpha$-structure $(H,T)$ is connected if $H$ is.
%

\begin{Property} \label{prop2SECchordal}
Let $H = (V,{\cal E})$ be an $\alpha$-acyclic hypergraph, and let $G$ be the 2-section graph $2SEC(H)$. 
Then $G$ is chordal and if moreover $H$ is a clutter then ${\cal E} = {\cal K}(G)$ (\textit{i.e.} the set  ${\cal E}$
of hyperedges of $H$ is equal to the set ${\cal K}(G)$ of maximal cliques of $G$).
\end{Property}

It follows that the number of hyperedges of a clutter is bounded by the number of its vertices since a chordal graph has at most $n$ maximal cliques.
The number of hyperedges of an $\alpha$-acyclic hypergraph which is not a clutter may be exponential in the number of vertices. 
% For instance, for any set $V$, the hypergraph $H = (V,P(V) \setminus \{\emptyset \})$ is $\alpha$-acyclic since $V$ is a hyperedge of $H$ (the tree whose edges are the pairs of hyperedges containing $V$ is a join tree of $H$).

A join tree of a connected $\alpha$-acyclic hypergraph $H = (V,{\cal E})$ can be defined from 
its \textit{weighted line graph}, where weights are defined as follows. The \textit{set associated with} 
a pair $\{X,Y\}$ of ${\cal E}$ is $X \cap Y$, and its \textit{weight}, denoted by $w(XY)$, is $|X \cap Y|$.
Let $K$ be a graph whose node set is ${\cal E}$. The weight of $K$
%
%, denoted by $w(K)$, 
%
is the sum 
of the weights of its edges. When considered as a weighted graph, 
$K$ is denoted by $K_w$. Thus $L_w(H)$ denotes the weighted line graph of $H$.

\begin{Characterization} \label{carMaxWeight}
\cite{abchapter} 
Let $H = (V,{\cal E})$ be an $\alpha$-acyclic (resp. connected $\alpha$-acyclic) hypergraph.
Then the join trees of $H$ are the maximum weight spanning trees of the weighted complete graph on ${\cal E}$ (resp. of $L_w(H)$).
\end{Characterization}

In particular the atom trees of a connected graph $G$ are the maximum weight spanning trees 
of the weighted intersection graph of the atoms of $G$, which is proved  in the case of  chordal graph in \cite{BP} (and extends to any connected graph through the chordal graph $G^+$ by Property~\ref{propG*}). 

%%%%%%%%%%%%%%%%%%%%%%%%%%%%%%%%%%%%%%%%%%%%
\section{Atom graphs} \label{sectAtomGraph}
%%%%%%%%%%%%%%%%%%%%%%%%%%%%%%%%%%%%%%%%%%%%

Atom graphs were used in \cite{InSilico} and formally introduced in \cite{HAL2010}.

\begin{Definition} \cite{HAL2010}
The {\em atom graph} of a graph $G$, denoted by {\em $AG(G)$}, is the graph $({\cal A}(G), E')$, 
where ${\cal A}(G)$ is the set of atoms and 
$E'$ the set of  pairs $\{A,B\}$ of ${\cal A}(G)$ such that $A \cap B$ is a clique minimal $(A \setminus (B \setminus A)$-separator of $G$.
\end{Definition}

\begin{figure}
\begin{center}
\begin{tikzpicture} 

\begin{scope} [scale = 0.6]
\node [ens] (A)  at (0,2) {$A$};
\node [ens] (B)  at (2,3) {$B$};
\node [ens] (C)  at (4,2) {$C$};
\node [ens] (D)  at (0,0) {$D$};
\node [ens] (E)  at (2,-1) {$E$};
\node [ens] (F)  at (4,0) {$F$};

\draw (A) -- (B)  node [midway,above left] {$\{1,2,3\}$}
                -- (C)  node [midway,above right] {$\{1,7\}$};
\draw (A) -- (D)  %node [midway,left] {$\{1\}$}
                -- (E) % node [midway,below left] {$\{1\}$}
                -- (F)  node [midway,below right] {$\{10,11\}$};
\draw (A) -- (E)  %node [midway,right] {$\{1\}$}
                -- (C)  %node [midway,below left] {$\{1\}$}
               -- (D) %node [midway,below] {$\{1\}$}
                -- (B)  %node [midway,below] {$\{1\}$}
                -- (E) ; %node [midway,right] {$\{1\}$}
\end{scope}

\end{tikzpicture}
\end{center} 

\caption{The atom graph of $G$ (the edge labels that are equal to $\{1\}$ are omitted).} \label{figAtomGraph}

\end{figure}
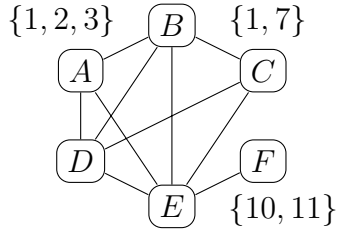

\begin{Example}
Figure~\ref{figAtomGraph}  shows the atom graph of the graph $G$ from Figure~\ref{figAtomTree}.
\end{Example}

%\begin{figure}[h!]
%\begin{center}
%\includegraphics[width=0.6\textwidth]{figAtomGraph.eps}
%\includegraphics[width=0.6\textwidth]{fig2}
%\end{center}
%\caption{The atom graph of $G$.}\label{figAtomGraph}
%\end{figure}

%%%%%%%%%%%%%%%%%%

%%%%%%%%%%%
In the definition of the atom graph, the word `clique' can be removed by Property~\ref{propInterAtoms} and the word `minimal' can be removed by Property~\ref{propDefAG}, which implies that for each pairs $\{A,B\}$ of ${\cal A}(G)$, each one of  
$A \setminus B$ and $B \setminus A$ is a subset of a full component of $A \cap B$.

\begin{Characterization} \label{carDefAG}
Let $G$ be a connected graph and let $A$ and $B$ be distinct atoms of $G$. Then $AB$ is an edge of $AG(G)$ if and only if $A \cap B$ is an $ab$-separator of $G$ for some $a \in A \setminus B$ and some $b \in B \setminus A$.
\end{Characterization}

\noindent
The following property immediately follows from Properties~\ref{cgatoms} and \ref{propG*Leimer}.

\begin{Property}\label{Prop:agisagcg}
 For a connected graph $G$, the atom graph of $G$ is the atom graph of the chordal graph $G^+$ 
%obtained by saturating the atoms of $G$. 
($G^+$ is defined in Notation~\ref{notGplus}).
\end{Property}

%\begin{Property}\cite{Habib1}\label{prop:cliquetreeedge}
 %Let $H$ be a chordal graph. Then each edge of the atom graph of $H$ belongs to at least one clique tree of $H$.
%\end{Property}

\noindent
Characterizations~\ref{carAGunionAT} and \ref{carATmaxAG} below give relationships between the atom graph and the atom trees. They are both
 proved for chordal graphs in \cite{Habib1} and also apply to any connected graph through the chordal graph $G^+$ by Properties~\ref{cgatoms}, \ref{propG*} and \ref{Prop:agisagcg}.

\begin{Characterization} \label{carAGunionAT}
 The atom graph of a connected graph $G$ is the union of all the atom trees of $G$.
\end{Characterization}

\begin{Characterization} \label{carATmaxAG}
The atom trees of a connected graph $G$ are the maximum weight spanning trees of the weighted atom graph of $G$.
\end{Characterization}
%
% 
% 
% In particular, we refind that the sum of the sizes of the maximal cliques of a chordal graph is bounded by $n+m$. 
% We also refind from the following corollory that the he sum of the sizes of its minimal separators is also bounded by $n+m$.

\noindent
To compute the edges of the atom graph from an atom tree, we will use the following characterization from \cite{Habib2} for chordal graphs, 
which also applies to any connected graph through the chordal graph $G^+$ by Properties~\ref{cgatoms}, \ref{propG*} and \ref{Prop:agisagcg}.

\begin{Characterization} \label{carEdgeAG}
Let $G$ be a connected graph, let $A$ and $B$ be distinct atoms of $G$ and let $T$ be an atom tree of $G$. Then
$AB$ is an edge of $AG(G)$ if and only if 
there is an edge $A'B'$ on the path $P_T(A,B)$ from $A$ to $B$ in the tree $T$ such that $A \cap B = A' \cap B'$.
\end{Characterization}

%%%%%%%%%%%%%%%%%%%%%%%%%%%%%%%%%%%%%%%%%%%%%%%%%%%%%%%%
\section{Computing the atom graph} \label{sectComputeAG}
%%%%%%%%%%%%%%%%%%%%%%%%%%%%%%%%%%%%%%%%%%%%%%%%%%%%%%%%

We know that given a connected graph $G$, an atom tree of $G$ (and therefore the atoms of $G$) can be computed in linear time if $G$ 
is chordal and in $O(min(n^{\alpha} \log n, nm,n(n+\overline{m})))$ time otherwise.
To compute the set of edges of the atom graph of $G$, a naive algorithm consists in computing for each pair $\{A,B\}$ of atoms 
of $G$ the connected components of $G(V \setminus (A \cap B))$ and determining whether $A \setminus B$ and $B \setminus A$ are in different components, which can be done in $O(m)$ time for each pair $\{A,B\}$ and therefore in $O(n^2m)$ time globally. 
\par
We will improve upon this to obtain a time which is no worse than that of computing an atom tree.
\par
Our first algorithm starts with an atom tree and the inclusion relation between the separators represented by the edges, 
and adds all the extra edges required to construct the atom graph. Our second algorithm 
starts with the weighted intersection graph of the atoms and repeatedly determines the edges of  weight 
$k$ which belong to the atom graph in decreasing order of $k$.
Both algorithms run in $O(n^2)$ time given these inputs. When only the graph is given as input, 
we obtain a complexity of $O(min(n^{\alpha} \log n, nm,n(n+\overline{m})))$ 
time, as will be detailed in this section.
\par
We introduce the following parameters which will be used in this section and in Section~\ref{sectComputeUJ}: 
$p$ denotes the number of atoms of $G$, $s$ the sum of their sizes and for each atom tree $T$ of $G$, $s_{\triangle}(T)$ 
denotes the sum of the symmetrical difference $X \triangle Y = (X \setminus Y) \cup (Y \setminus X)$ for each edge $XY$ of $T$. 

\begin{Notation}
For each connected graph $G$,
$p = |{\cal A}(G)|$, $s = \Sigma_{X \in {\cal A}(G)} |X|$, and for each 
atom tree $T = ({\cal A}(G),E_T)$ of $G$ $s_{\triangle}(T) = \Sigma_{XY \in {E_T}} |X \triangle Y|$.
\end{Notation}

Note that $p \leq n$ since $G$ has at most $n$ atoms and that $s \leq n+m$ since the sum of the sizes of the atoms of $G$ 
is bounded by $n+m$ by   Property~\ref{propSomAtoms}.
The parameters $p$, $s$ and $s_{\triangle}(T)$ are introduced for two reasons. 
First, they will be used to extend the complexity results of this section to the context of $\alpha$-acyclic hypergraphs 
in Section~\ref{sectComputeUJ} with appropriate extensions of the definitions of these parameters. 
Second, it can lead to a better complexity for graph classes for which these parameters have specific bounds.
\par

As will be detailed in Section 4.1, we will also need the edge-inclusion relation \textit{sub}, which for an atom 
tree tests for inclusion the separators represented by two edges.

\begin{Definition} \label{defsub}
Let $T = ({\cal A},E_T)$  be tree, with ${\cal A}$ a subset of $P(V)$ for some set $V$.
We call  {\em subset relation} of $T$ the relation $sub$ in $E_T$ defined by : $\forall XY, X'Y' \in E_T$ $sub(XY,X'Y') \Leftrightarrow X \cap Y \subseteq X' \cap Y'$.
\end{Definition}

We will show in Sections 4.1 and 4.2 the following complexity result:

\begin{Theorem} \label{thComplexAG}
The atom graph of a connected graph $G$ can be computed : \\
a) in $O(n^2)$ time from either an atom tree of $G$ and its subset relation or the weighted intersection graph of the atoms of $G$,  \\
b) in $O(min(n^{\alpha},nm,n(n+\overline{m^+})))$ time  from an atom tree of $G$, \\
c) in $O(min (n^{\alpha},nm))$ time  from the set of atoms of $G$, \\
d) in $O(min(n^{\alpha} \log n, nm,n(n+\overline{m})))$ time from $G$, \\
where $\overline{m^+}$ denotes the number of edges of $\overline{G^+}$ ($G^+$ is defined in Notation~\ref{notGplus}). 
\end{Theorem}

For a chordal graph $H$, the atom graph can thus be computed in 
$O(min (n^{\alpha},nm,n(n+\overline{m})))$ time, 
since in that case $G^+ = G$ and an atom tree (clique tree) of $G$ be computed in linear time.
\par
Other approaches are possible, but with no improvement of the time complexity. For instance, as the atom trees of a graph $G$ are obtained from the atom trees of a minimal triangulation $H$ of $G$ by merging the maximal cliques of $H$ into the atoms of $G$ \cite{AtomTree}, the atom graph of $G$ is obtained from the atom graph of $H$ by merging  the same maximal cliques of $H$.
\par
The different items of Theorem~\ref{thComplexAG} are detailed in the following results: item a) follows from Theorem~\ref{thAGsub} and Corollary~\ref{corAGmax},
iten b) follows from item c) and Theorem~\ref{thAGdelta},
item c) follows from item a) and Proposition~\ref{propAGmax},
and iten d) follows from item b) and from the fact that an atom tree of $G$ can be computed in $O(min(n^{\alpha} \log n, nm,n(n+\overline{m})))$ time.
%§§§§§§§§§§§§§§§§§§§§§§§§§§§§§§
\subsection{Algorithm Forest Join}
%§§§§§§§§§§§§§§§§§§§§§§§§§§§§§§
Our first algorithm, Forest Join, is based on Characterization \ref{carAGsub}.  Given an atom tree $T$, a \ms S is represented by one or several 
edges of $T$.  If we remove these edges, we obtain a forest.  Let us now furthermore shrink this forest by removing the nodes which do not contain $S$. 
Any edge between two nodes of different trees of the resulting forest will correspond to an edge of the atom graph which also represents S, 
and all the $S$ representatives are thus encountered.
\par
To implement this remarkable property, our algorithm processes the edges of the atom tree one by one, and computes the relevent nodes and edges with the help of relation \textit{sub}.

\begin{Characterization} \label{carAGsub}
Let $G$ be a connected graph, let $T$ be an atom tree of $G$  and let $S$ be a minimal separator of $G$. 
Then the edges of $AG(G)$ associated with $S$ are the pairs of nodes of $T$ whose endpoints are in different connected components of $T({\cal A}_S)-E_S$, where ${\cal A}_S$ is the set of nodes of $T$ containing $S$ and  $E_S$ is the set of edges of $T$ associated with $S$.
\end{Characterization}

\begin{Proof}
Let $\{X,Y\}$ be a pair of nodes of $T$.
Let us show that $XY$ is an edge of $AG(G)$ associated with $S$ if and only if $X$ and $Y$ are in different connected components of $T({\cal A}_S)-E_S$, i.e. 
by Characterization~\ref{carEdgeAG} and the fact that $T({\cal A}_S)$ is connected, that
there is an edge $X'Y'$ of $P_T(X,Y)$ such that $S =X \cap Y = X' \cap Y'$ if and only if $P_T(X,Y)$ is a path in $T({\cal A}_S)$ having an edge $X'Y'$ in $E_S$. \\
$\Rightarrow$: as $S =X \cap Y$ $P_T(X,Y)$ is a path in $T({\cal A}_S)$, and as $S = X' \cap Y'$, $X'Y'$ is in $E_S$. \\
$\Leftarrow$: as $X$ and $Y$ are in ${\cal A}_S$, $S \subseteq X \cap Y$.
Hence $S \subseteq X \cap Y \subseteq X' \cap Y' = S$, and therefore $S = X \cap Y = X' \cap Y'$.
\end{Proof}

Algorithm Forest Join computes the edges of the atom graph of $G$ according to Characterization~\ref{carAGsub}. 
Given an atom tree $T$ of $G$ and its subset relation $sub$, it  scans the edges of $T$ and for 
each edge $AB$, it computes the set of edges of the atom graph associated with the minimal separator $S$ 
associated with $AB$ if it has not be computed yet,\textit{ i.e.} if $AB$ does not belong to the set of 
edges computed so far.
\par

 It calls Algorithm Components which computes the connected components of the forest $T({\cal A}_S)-E_S$ defined in 
 Characterization~\ref{carAGsub}. The relation $sub$ enables us to compute these components at no extra cost than a simple 
 tree search: $T({\cal A}_S)$ is the subtree of $T$ whose edges $XY$ are associated with supersets of $S$, \textit{i.e.} 
 satisfy $sub(AB,XY)$, and $E_S$ is the set of edges  $XY$ of $T$  associated with $S$, i.e. 
 such that $sub(AB,XY)$ and $sub(XY,AB)$. 
 \par
 
 In Algorithm Components, $k$ is the current number of connected components, 
$C_1, \ldots, C_k$ are the current components, $Queue$ contains the nodes of $T$ that are reached 
but not processed yet and for each reached node $X$, $numComp(X)$ is the index $i$ of the 
component $C_i$ containing $X$ and $pred(X)$ is the node of $T$ it has been reached from 
(and which should not be processed again).
\begin{algorithm}[h!] 
\SetKwInOut{Input}{input}
\SetKwInOut{Output}{output}
\textbf{Algorithm Forest Join}\vfill
\Input{An atom tree $T = ({\cal A}, E_T)$ of a connected graph $G$ and its subset relation $sub$.}
\Output{The atom graph of $G$.}
%\vspace{-0,1cm}
\BlankLine 
{ % begin 
$E' \leftarrow \emptyset$\;
\ForEach {$AB \in E_T$}
{% foreach
\If {$AB \notin E'$}
{% if
// the edges associated with $A \cap B$ are not in $E'$ yet \\
$CompSet \leftarrow$ \textbf{Components}$(T, AB,sub)$\;
\ForEach {$\{C,C'\} \subseteq CompSet$}
{% foreach2
\ForEach {$X \in C$}
{% foreach3
\ForEach {$Y \in C'$}
{% foreach4
Add $XY$ to $E'$\;
}% foreach4
}% foreach3
}% foreach2
}% if
}% foreach
return $({\cal A},E')$\;
} % begin
\end{algorithm}

\par

\begin{algorithm}[h!] 
\SetKwInOut{Input}{input}
\SetKwInOut{Output}{output}
\textbf{Algorithm Components}\vfill
\Input{An atom tree $T$ of an connected graph, an edge $AB$ of $T$ and the subset relation $sub$ of $T$.}
\Output{The set of connected component of $T({\cal A}_S)-E_S$, where $S = A \cap B$, ${\cal A}_S$ is the set of nodes of $T$ containing $S$ and  $E_S$ is the set of edges of $T$ associated with $S$.}
%\vspace{-0,1cm}
\BlankLine 
{ % begin 
$k \leftarrow 1$; $C_1 \leftarrow \{A\}$; $numComp(A) \leftarrow 1$; $Queue \leftarrow \{A\}$\;
\While{$Queue \neq \emptyset$}
{% while
Remove a node $X$ from $Queue$\;
\ForEach {$Y \in N_T(X)$}
{% foreach
\If {$(Y \neq pred(X)) \wedge sub(AB,XY)$}
{% if
\If {$sub(XY,AB)$}
{% if2
// $XY$ associated with $S$, begin a new component \\
$k \leftarrow k +1$; $C_k \leftarrow \emptyset$; $i \leftarrow k$\;
}% if2
\Else
{% else
$i \leftarrow numComp(X)$\; 
}% else
Add $Y$ to $C_i$; $numComp(Y) \leftarrow i$\; 
$pred(Y) \leftarrow X$; Add $Y$ to $Queue$\;
}% if
}% foreach
}% while
return $\{C_1, \ldots , C_k\}$\;
} % begin
\end{algorithm}

\begin{figure}
\begin{center}
\begin{tikzpicture} [scale = 0.6]

\begin{scope}%[scale = 0.6]

\node [ens] (A)  at (0,2) {$A$};
\node [ens] (B)  at (2,3) {$B$};
\node [ens] (C)  at (4,2) {$C$};
\node [ens] (D)  at (0,0) {$D$};
\node [ens] (E)  at (2,-1) {$E$};
\node [ens] (F)  at (4,0) {$F$};

\draw (A) -- (B)  node [midway,above left] {$\{1,2,3\}$}
                -- (C)  node [midway,above right] {$\{1,7\}$};
\draw (A) -- (D)  node [midway,left] {$\{1\}$}
                -- (E)  node [midway,below left] {$\{1\}$}
                -- (F)  node [midway,below right] {$\{10,11\}$};

\draw  (2,-2.5) node {$T$};
\end{scope}

\begin{scope}[xshift = 200]
%[scale = 0.6,xshift = 200]

\node [ens] (A)  at (0,2) {$A$};
\node [ens] (B)  at (2,3) {$B$};
\node [ens] (C)  at (4,2) {$C$};
\node [ens] (D)  at (0,0) {$D$};
\node [ens] (E)  at (2,-1) {$E$};
%\node [ens] (F)  at (4,0) {$F$};

\draw (A) -- (B) 
                -- (C)  ;
\draw [dot] (A) -- (D)  
                -- (E) ;
\draw [dot] (A) -- (E) 
                -- (C) 
               -- (D) 
                -- (B) 
                -- (E) ; 
\draw  (2,-2.5) node {$S = \{1\}$};
\end{scope}

\begin{scope}[xshift = 400]
%[scale = 0.6,xshift = 400]

\node [ens] (A)  at (0,2) {$A$};
\node [ens] (B)  at (2,3) {$B$};
\node [ens] (C)  at (4,2) {$C$};
\node [ens] (D)  at (0,0) {$D$};
\node [ens] (E)  at (2,-1) {$E$};
\node [ens] (F)  at (4,0) {$F$};

\draw (A) -- (B) 
                -- (C)  ;
\draw (A) -- (D)  
                -- (E) 
                -- (F) ;
\draw (A) -- (E) 
                -- (C) 
               -- (D) 
                -- (B) 
                -- (E) ; 
\draw  (2,-2.5) node {$AG(G)$};
\end{scope}

\end{tikzpicture}
\end{center} 

\caption{An execution of Algorithm Forest Join.} \label{figAGsub}

\end{figure}
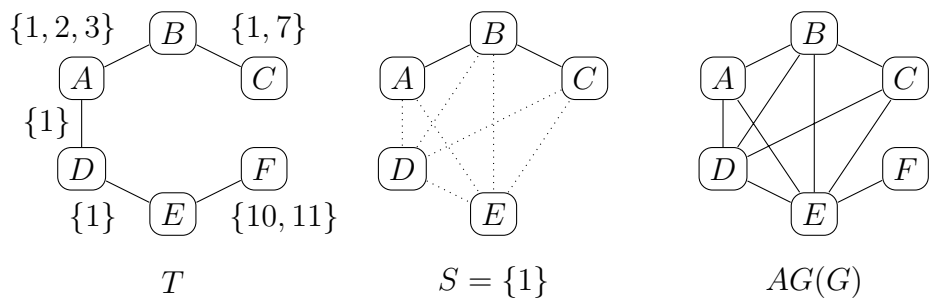

\begin{Example} \label{exAGsub}
% Figure 4
Figure~\ref{figAGsub} shows an atom tree $T$ of the graph $G$ from Figure~\ref{figAtomTree} and an execution of 
Algorithm Forest Join on $T$ and its subset relation. It shows the forest $T({\cal A}_S)-E_S)$ for $S=\{1\}$, 
where the edges of the atom graph associated with $S$ 
are represented by dotted lines.
For each clique \ms $S$ different from  $\{1\}$, as ${\cal A}_S$ is of size $2$, $AG(G)$ has a unique edge 
associated with $S$ which is also an edge of $T$. So $AG(G)$ is obtained from $T$ by adding the edges 
associated with $\{1\}$ that are not already present in $T$.
\end{Example} 

%\begin{figure}[h!] 
% Figure 4
%\begin{center}
%\includegraphics[width=0.6\textwidth]{figFJ}
%\end{center}
%\caption{An execution of Algorithm Forest Join.} \label{figAGsub}
%\end{figure}

%%%%%%%%%%%%%%%

%%%%%%%%%%%%%%
%
\begin{Theorem} \label{thAGsub}
Given an atom tree of a connected graph $G$ and its subset relation,
 Algorithm Forest Join computes the atom graph of $G$ in $O(p^2)$
 time, and therefore in $O(n^2)$ time.
\end{Theorem}

\begin{Proof}
The correctness follows from Characterization~\ref{carAGsub}. 
Let us prove the time complexity.
As Algorithm Components runs in $O(p)$ time and is called less than $p$ times, it globally costs $O(p^2)$ time.
As an edge $XY$ is added to $E'$ at most once (when processing the first edge of $T$ associated with $X \cap Y$), the number of edge additions to $E'$ is bounded by $p^2$.
 Hence Algorithm Forest Join runs in $O(p^2)$ time, and therefore in $O(n^2)$ time since $G$ has at most $n$ atoms ($p \leq n$).
\end{Proof}

% Now we want to evaluate the time complexity when the input is the graph $G$ itself.

To evaluate the time complexity of computing the atom graph of $G$ from an atom tree $T$ of $G$ using Algorithm Forest Join, 
we need the time complexity of computing the subset relation of $T$.

\begin{Proposition} \label{propAGsub}
Given an atom tree of a connected graph, its subset relation  can be computed in $O(min (n^{\alpha},ps))$ time, and therefore
in $O(min (n^{\alpha},nm))$ time.
\end{Proposition}

\begin{Proof}
It follows from the proof of Property~\ref{propSomMinSep} that the sets $X \cap Y$ for each edge of $T$ can be computed in $O(s)$ time.
and that the sum of their sizes is bounded by $s$. 
As the inclusion of $X \cap Y$ in $X' \cap Y'$ or not can be determined in $O(|X \cap Y|)$ time, the subset relation can be computed in $O(s +ps)$ time, i.e. in $O(ps)$ time. \\
Alternatively, as $sub(XY,X'Y')$ is equivalent to $|X \cap Y| = |(X \cap Y) \cap (X' \cap Y')|$, it can be evaluated in $O(1)$ time, and therefore in $O(p^2)$ time globally, provided that the values of $|X \cap Y|$ and $|(X \cap Y) \cap (X' \cap Y')|$ have been pre-computed.
The values of $X \cap Y$ and $|X \cap Y|$ for each edge $XY$ of $T$ can be computed in $O(s)$ time, and the values of $|(X \cap Y) \cap (X' \cap Y')|$ in $O((n+p)^{\alpha})$ time since they are the elements of the product of the transpose of $M$ by $M$, where $M$ is the $n \times (p-1)$ incidence matrix of the (possibly non-simple) hypergraph whose vertex set is $V$ and whose hyperedges are the sets $X \cap Y$ for each edge $XY$ of $T$.
Hence this alternative complexity is in $O(p^2 + s +(n+p)^{\alpha})$ time, i.e. in $O((n+p)^{\alpha})$ time since $p^2 \leq (n+p)^2$, $s \leq np \leq (n+p)^2$ and $2 \leq \alpha$. \\ 
We obtain a complexity in $O(min((n+p)^{\alpha},ps))$ time, i.e. in $O(min (n^{\alpha},ps))$ time since $p \leq n$ and therefore in  $O(min (n^{\alpha},nm))$ time since $s \leq n+m$ by   Property~\ref{propSomAtoms}
\end{Proof}

It follows that the atom graph can be computed from an atom tree in $O(min (n^{\alpha},nm))$ time.
\par

We will now discuss using the 2-pairs of the graph $G^+$ defined in Notation~\ref{notGplus} to obtain 
an alternative complexity in $O(n(n+\overline{m^+}))$ time, where $\overline{m^+}$ is the number 
of edges of the complement of $G^+$, through the following lemma.

\begin{Lemma} \label{lemAGdelta}
Let $T$ be atom tree of a connected graph. Then $s_{\triangle}(T) \leq n+ t$, where $t$ is the number of 2-pairs of $G^+$.
\end{Lemma}

\begin{Proof}
We consider a rooted directed tree $T_r = ({\cal A}(G), U)$  obtained from $T$ by choosing an arbitrary root.
Thus $\Sigma_{XY \in E_T} |X \triangle Y| = \Sigma_{(X,Y) \in U} |X \triangle Y|$.
$\Sigma_{(X,Y) \in U} |Y \setminus X| \leq n$ since each vertex $x$ of $G$ belongs to $Y \setminus X$ for at most one edge of $T_r$, namely the edge $(X,Y)$ such that $Y$ is the root of the subtree of $T_r$ induced by the nodes containing $x$.
Hence it is sufficient to show that 
$\Sigma_{(X,Y) \in U} |X \setminus Y| \leq t$.
It is shown in \cite{AtomTree} that if $G$ is chordal then this sum is bounded by the number of 2-pairs of $G$. So it is bounded by $t$ since $G^+$ is chordal and by Property~\ref{propG*} $T$ is also an atom tree of $G^+$.
\end{Proof}

\begin{Theorem} \label{thAGdelta}
The atom graph of a connected graph $G$ can be computed from an atom tree $T$ of $G$
 in $O(p(n+s_{\triangle}(T)))$ time, and therefore in $O(n(n+\overline{m^+}))$ time, where $\overline{m^+}$ is the number of edges of the complement of $G^+$.
\end{Theorem}

\begin{Proof}
We consider the variant of Algorithm Forest Join where the subset relation $sub$ is not given as an input and ''$sub(AB,XY)$'' and ''$sub(XY,AB)$'' in Algorithm Components are replaced as follows. 
Condition $sub(XY,AB)$ can be replaced by $|A \cap B| = |X \cap Y|$ since in the algorithm $XY$ satisfies $sub(AB,XY)$.
The values of $|X \cap Y|$ for each edge $XY$ of $T$ can be pre-computed in $O(np)$ time.
Let $S = A \cap B$.
As $S$ is a subset of $X$ in the algorithm, condition $sub(AB,XY)$ is equivalent to $(X \setminus Y) \cap S = \emptyset$, which can be evaluated in $O(|X \setminus Y |)$ time, provided that the sets $X \cap Y$, $X \setminus Y$ and $Y \setminus X$ for each edge $XY$ of $T$ have been pre-computed, which can be done in $O(np)$ time.
Thus we  add to the time complexity of Algorithm Forest Join in $O(p^2)$
a pre-computation time in $O(np)$ and $O(s_{\triangle}(T))$ time per call to Components. We obtain a time complexity in $O(p^2 + np+p*s_{\triangle}(T))$, and therefore in $O(p(n+s_{\triangle}(T)))$ since $p-1 \leq s_{\triangle}(T)$ (because the nodes of $T$ are pairwise distinct).
We conclude with Lemma~\ref{lemAGdelta}
\end{Proof}

The 2-pairs of a chordal graph are closely related to its atom graph.

\begin{Characterization} \label{carAGdelta}
Let $G$ be a connected chordal graph, and let $\{x,y\}$ be a pair of vertices of $G$.
Then $xy$ is a 2-pair of $G$ if and only if there is an edge $KL$ of $AG(G)$ such that $x \in K \setminus L$ and $y \in L \setminus K$.
\end{Characterization}

\begin{Proof}
$\Rightarrow$: let $S = N(x) \cap N(y)$.
As $S$ is a minimal separator of $G$ and $G$ is chordal, $S$ is a clique. Let $K$ (resp. $L$) be a maximal clique containing $\{x\} \cup S$ (resp. $\{y\} \cup S$). 
$S \subseteq K \cap L \subseteq (N(x) \cup \{x\}) \cap (N(y) \cup \{y\}) = N(x) \cap N(y) = S$.
Hence $S = K \cap L$, and therefore $KL$ is an edge  of $AG(G)$ with $x \in K \setminus L$ and $y \in L \setminus K$. \\
$\Leftarrow$: let $S = K \cap L$.
As $KL$ is an edge of $AG(G)$, $S$ is a minimal $xy$-separator. 
As $G$ is chordal $K$ and $L$ are cliques, so $S \subseteq N(x) \cap N(y)$, and as $S$ is an $xy$-separator $N(x) \cap N(y) \subseteq S$. 
Hence $S = N(x) \cap N(y)$, and therefore$xy$ is a 2-pair of $G$.
\end{Proof}

It follows that the number of 2-pairs of a connected chordal graph $G$ is bounded by the sum of the products $|K \setminus L|*|L \setminus K|$ for each edge $KL$ of its atom graph. In particular, in a graph class (of non-necessarily chordal graphs) in which the values of $|A \setminus B|$ (and $|B \setminus A|$) for each edge $AB$ of the atom graph are bounded by a given constant, for instance if the sizes of the atoms are bounded by a constant, the atom graph can be computed from an atom tree
 in $O(n(n+m'))$ time where $m'$ is the number of edges of the computed atom graph.
The number of 2-pairs  is not equal in general to the sum of the products $|K \setminus L|*|L \setminus K|$ for each edge $KL$ of its atom graph since a 2-pair may be associated with several edges of the atom graph. Considering the same relation between the 2-pairs and the edges of an atom tree $T$ of $G$, a pair $\{x,y\}$ associated with an edge $KL$ of $T$ is a 2-pair since $KL$ is an edge of the atom graph, but the converse does not hold. Contrary to the atom graph, $\{x,y\}$ can be associated with at most one edge of $T$, namely the unique edge connecting the subtrees of $T$ induced by the sets of nodes containing $x$ and $y$ respectively, which are necessarily disjoint and at distance $1$ from each other in $T$.

\par
This alternative time complexity in $O(n(n+\overline{m^+}))$ considers the worst case where Algorithm Components searches the whole tree $T$, whereas it only searches the set ${\cal A}_S$ and its neighborhood which may be very small w.r.t. the set of nodes of $T$. For the same reason, it may be more efficient in practice to execute Algorithm Forest Join without pre-computing the subset relation $sub$ and directly evaluate $sub(AB,XY)$ and $sub(XY,AB)$ when needed.

%§§§§§§§§§§§§§§§§§§§§§§§§§§§§§§§§§§§§§§
\subsection{Algorithm AG-max-weight}
%§§§§§§§§§§§§§§§§§§§§§§§§§§§§§§§§§§§§§§
Our second algoirithm, AG-max-weight, takes as input the weighted intersection graph of 
the atoms (which, in the case of a chordal graph, is the clique graph) and repeatedly adds 
% all the possible maximal weight edges which do not violate a tree structure.
the edges of weight $k$ in decreasing order of $k$.

By Characterization~\ref{carMaxWeight} the atom trees of a connected graph $G$ are the maximum weight spanning trees 
of the weighted intersection graph of the atoms of $G$. We will present a general algorithm computing the union of 
the maximum weight spanning trees of $G_w$ for each weighted connected  graph $G_w$ with natural integer weights on the
edges. 
This general algorithm called Union-max-weight is inspired from
 the following algorithm from Kruskal which computes a \textit{minimum} weight spanning tree of $G_w$ : 
 initialize graph $T'$ 
 as edgeless and for each edge $xy$ of $G_w$ in \textit{increasing} order of weight, add $xy$ to $T'$ if and 
 only if $x$ and $y$ are in different connected components of $T'$. 
 As we want to compute a \textit{maximum} weight spanning tree, we will process 
the edges in \textit{decreasing} order of weight; 
 the algorithm computes each maximum weight spanning tree of $G_w$.
Thus an edge $xy$ of weight $k$ may be added to $T'$ by this last algorithm if and only if $x$ and $y$ are in different 
connected components of $T'$ just after processing the edges of weight at least $k+1$. These components are independent 
from the graph $T'$ computed so far by item a) of Lemma~\ref{lemUnionmax} below. 

\begin{Lemma} \label{lemUnionmax}
Let $G_w = (V,E,w)$ be a weighted connected graph  with natural integer weights on the edges, let $T$ be a maximum weight spanning tree of $G$, 
let $UM$ be the union of the maximum weight spanning trees of $G$ and for a natural integer $k$, let $G_k$ (resp. $T_k$, $UM_k$) be the graph whose vertex set is $V$ and whose edges are the edges of $G$ (resp. $T$, $UM$) of weight at least $k$. 
Then  \\
a) $G_k$, $T_k$ and $UM_k$ have the same connected components, \\
b) the edges of $UM$ of weight $k$ are the edges of $G$ of weight $k$ whose endpoints are in different connected components of $UM_{k+1}$.
\end{Lemma}

\begin{Proof}
a) As each connected component of $T_k$ is a subset of a connected component of $UM_k$ which is itself a subset of a connected component of $G_k$, it is sufficient to show that each connected component of $G_k$ is a subset of a connected component of $T_k$, or equivalently that for each edge $xy$ of $G_k$, $P_T(x,y)$ is a path in $T_k$. Let $xy$ be an edge of $G_k$. For each edge $x'y'$ of $P_T(x,y)$ $w(xy) \leq w(x'y')$ (otherwise $(T-\{x'y'\} )+\{xy\}$ would be a spanning tree of $G$ of strictly greater weight than $T$), so $P_T(x,y)$ is a path in $T_k$. \\
b) Let $xy$ be an edge of $G$ of weight $k$. Let us show that $xy$ is an edge of $UM$ if and only if $x$ and $y$ are in different connected components of $UM_{k+1}$. We assume that $xy$ is an edge of $UM$. Let $T$ be a maximum weight spanning tree of $G$ such that $xy$ is an edge of $T$. $x$ and $y$ are in different connected components of $T_{k+1}$, and therefore of $UM_{k+1}$ by a). Conversely we assume that $x$ and $y$ are in different connected components of $UM_{k+1}$. Let $T$ be a maximum weight spanning tree of $G$. As $x$ and $y$ are in different connected components of $T_{k+1}$, there is an edge $x'y'$ of $P_T(x,y)$ of weight at most $k$. Then $(T-\{x'y'\} )+\{xy\}$ is a maximum weight spanning tree of $G$, and therefore $xy$ is an edge of $UM$.
\end{Proof}

Item b) of Lemma~\ref{lemUnionmax} provides an inductive definition of the edges of weight $k$ of the union of the maximum weight spanning trees, and therefore a simple iterative algorithm to compute them.
Thus Algorithm Union-max-weight computes the union of the maximum weight spanning trees of $G$ by initializing a set $F$ with the empty set and adding to $F$, for each weight value $k$ in decreasing order, the edges $xy$ of $G$ of weight $k$ such that $x$ and $y$ are in different connected components of the graph $(V,F)$ in its state just after adding the edges of weight strictly grater than $k$.
\par
In Algorithm Union-max-weight, $k$ is the current value of weight, the sets $C_i$ are the connected components of the graph $(V,F)$ in its state at the beginning of iteration $k$ and for each vertex $x$, $numComp(x)$ is the index $i$ of the component $C_i$ containing $x$.
The algorithm is similar to the ``maximum weight'' variant of Kruskal's algorithm, the difference being that Kruskal's algorithm considers the connected components of the graph (tree) $(V,F)$ being computed in its current state instead of in its state at the beginning of iteration $k$, and therefore would update the variables $C_i$ and $numCom$ just after each addition of an edge to $F$. It follows that the algorithms and complexity results already published on Kruskal's algorithm hold for the computation of the union of the maximum weight spanning trees. In particular the complexity can be improved by using a sophisticated UNION-FIND data structure. However, the simple algorithm presented here is sufficient to compute the atom graph in $O(n^2)$ time.

\begin{algorithm}[h!] 
\SetKwInOut{Input}{input}
\SetKwInOut{Output}{output}
\textbf{Algorithm AG-max-weight}\vfill
\Input{The weighted intersection graph $IG_w$ of the atoms of a connected graph $G$.}
\Output{The atom graph of $G$.}
%\vspace{-0,1cm}
\BlankLine 
{ % begin 
return \textbf{Union-max-weight}$(IG_w)$\;
} % begin
\end{algorithm}

\begin{algorithm}[h!] 
\SetKwInOut{Input}{input}
\SetKwInOut{Output}{output}
\textbf{Algorithm Union-max-weight}\vfill
\Input{A weighted connected graph $G_w = (V,E,w)$, with natural integer weights on the edges}
\Output{The union of the maximum weight spanning trees of $G_w$.}
%\vspace{-0,1cm}
\BlankLine 
{ % begin 
Compute the maximum weight $w_{max}$ of an edge of $G_w$ and for each $k$ in $[1,w_{max}]$ the set $E_k$ of edges of $G_w$ of weight $k$\;
$i \leftarrow 0$\;
\ForEach {$x \in V$}
{% foreach
$i \leftarrow i+1$; $C_i \leftarrow \{x\}$; $numComp(x) \leftarrow i$\;
}% foreach
$F \leftarrow \emptyset$\;
\ForEach {$k=w_{max}$ downto $0$} 
{% foreach
\ForEach {$xy \in E_k$} 
{% foreach2
\If {$numComp(x) \neq numComp(y)$} 
{% if
Add $xy$ to $F$\;
}% if
}% foreach2
\ForEach {$xy \in E_k$} 
{% foreach2
\If {$numComp(x) \neq numComp(y)$} 
{% if
$i \leftarrow numComp(x)$; $j \leftarrow numComp(y)$; $C_i \leftarrow C_i \cup C_j$\;
\ForEach {$z \in C_j$} 
{% foreach3
$numComp(z) \leftarrow i$\;
}% foreach3
}% if
}% foreach2
}% foreach
return $(V,F)$\;
} % begin
\end{algorithm}

\begin{figure}
\begin{center}
\begin{tikzpicture} [scale = 0.6]

\begin{scope}%[scale = 0.6]

\node [ens] (A)  at (0,2) {$A$};
\node [ens] (B)  at (2,3) {$B$};
\node [ens] (C)  at (4,2) {$C$};
\node [ens] (D)  at (0,0) {$D$};
\node [ens] (E)  at (2,-1) {$E$};
\node [ens] (F)  at (4,0) {$F$};

\draw (A) -- (B)  node [midway,above left] {$3$}
                -- (C)  node [midway,above right] {$2$}
                -- (A) ; %node [midway,left] {$1$}
\draw (A) -- (D)  %node [midway,left] {$1$}
                -- (E)  %node [midway,below left] {$1$}
                -- (F)  node [midway,below right] {$2$};
\draw (A) -- (E) 
                -- (C) 
               -- (D) 
                -- (B) 
                -- (E) ; 
\end{scope}

\begin{scope}[xshift = 200]
%[scale = 0.6,xshift = 200]

\node [ens] (A)  at (0,2) {$A$};
\node [ens] (B)  at (2,3) {$B$};
\node [ens] (C)  at (4,2) {$C$};
\node [ens] (D)  at (0,0) {$D$};
\node [ens] (E)  at (2,-1) {$E$};
\node [ens] (F)  at (4,0) {$F$};

\draw (A) -- (B)  node [midway,above left] {$3$}
                -- (C)  node [midway,above right] {$2$};
\draw (E) -- (F)  node [midway,below right] {$2$};

\end{scope}

\begin{scope}[xshift = 400]
%[scale = 0.6,xshift = 400]

\node [ens] (A)  at (0,2) {$A$};
\node [ens] (B)  at (2,3) {$B$};
\node [ens] (C)  at (4,2) {$C$};
\node [ens] (D)  at (0,0) {$D$};
\node [ens] (E)  at (2,-1) {$E$};
\node [ens] (F)  at (4,0) {$F$};

\draw (A) -- (B)  node [midway,above left] {$3$}
                -- (C)  node [midway,above right] {$2$};
\draw (A) -- (D)  %node [midway,left] {$1$}
                -- (E)  %node [midway,below left] {$1$}
                -- (F)  node [midway,below right] {$2$};
\draw (A) -- (E) 
                -- (C) 
               -- (D) 
                -- (B) 
                -- (E) ; 
\end{scope}
\end{tikzpicture}
\end{center} 

\caption{An execution of Algorithm AG-max-weight (the edge labels that are equal to $1$ are omitted).} \label{figAGmax}
\end{figure}
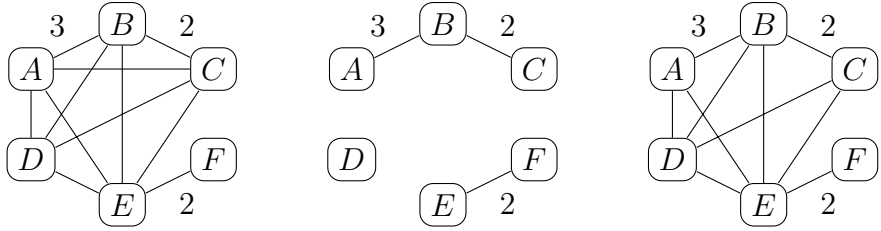

\begin{Example} \label{exAGmax}
% Figure 5
Figure~\ref{figAGmax} shows  the weighted intersection graph of the atoms of the graph $G$ from Figure~\ref{figAtomTree} and
an execution of Algorithm AG-max-weight, i.e. Algorithm Union-max-weight, on this weighted graph.
It shows the state of the computed graph before and after adding the edges of weight $1$.
\end{Example}

%\begin{figure}[h!] 
% Figure 4
%\begin{center}
%\includegraphics[width=0.6\textwidth]{figMW}
%\end{center}
%\caption{An execution of Algorithm AG-max-weight (the edge labels that are equal to $1$ are omitted).} \label{figAGmax}
%\end{figure}

%%%%%%%%%

%%%%%%%%%%%%%%

\begin{Theorem} \label{thUnionmax}
Given a weighted connected graph $G_w = (V,E,w)$ with natural integer weights on the edges,
 Algorithm Union-max-weight computes the union of the maximum weight spanning trees of $G_w$ in $O(w_{max}+n^2)$ time, where $w_{max}$ is the maximum weight  of an edge of $G_w$.
\end{Theorem}

\begin{Proof}
It follows from Lemma~\ref{lemUnionmax} that the property $P$ defined below is an invariant of the main foreach loop, using the notation $UM_k$ of this lemma, \\
$P$ : $UM_k = (V,F)$ and $\forall x \in V$ ($C_{numComp(x)}$ is the connected component of $UM_k$ containing $x \wedge \forall y \in C_{numComp(x)}$ $numComp(x) = numComp(y)$), \\
which proves the correctness of the algorithm. \\
Let us prove its time complexity.
$w_{max}$ and the sets $E_k$ can be computed and scanned in the internal foreach loops in $O(w_{max}+ m)$ time by storing the elements of $E_k$ at index $k$ of an array.
The first internal foreach loop runs in $O(m)$ time globally, and the second one in $O(n^2)$ time globally since merging two connected components is in $O(n)$ time and is performed $n-1$ times. Hence the algorithm runs in $O(w_{max}+n^2)$ time.
\end{Proof}

\begin{Corollary} \label{corAGmax}
Given the weighted intersection graph of the atoms of a connected graph $G$,
 Algorithm AG-max-weight computes the atom graph of $G$ in $O(n+p^2)$ time, and therefore in $O(n^2)$ time.
\end{Corollary}

To evaluate the time complexity of computing the atom graph of $G$ from the set of atoms of $G$ using Algorithm AG-max-weight, we need the time complexity of computing the the weighted intersection graph of the atoms of  $G$.

\begin{Proposition} \label{propAGmax}
Given the set of atoms of a connected graph $G$, the weighted intersection graph of the atoms of $G$ can be computed in $O(min (n^{\alpha},ps))$ time, and therefore in $O(min (n^{\alpha},nm))$ time.
\end{Proposition}

\begin{Proof}
As $|X \cap Y|$ can be computed in $O(|Y|)$ time, computing 
$|X \cap Y|$ for each pair $\{X,Y\}$ of atoms of $G$ is in $O(ps)$ time.
Alternatively these values can be computed in $O((n+p)^{\alpha})$ time since they are the elements of the product of the transpose of $M$ by $M$, where $M$ is the $n \times p$ incidence matrix of the hypergraph $(V,{\cal A}(G))$ (which will be called the atom hypergraph of $G$ in Section~\ref{sectAtomHypergraph}), i.e. in $O(n^{\alpha})$ time since $p \leq n$.
We obtain a time complexity  in $O(min (n^{\alpha},ps))$, and therefore in $O(min (n^{\alpha},nm))$ since $s \leq n+m$ by   Property~\ref{propSomAtoms}.
\end{Proof}

It follows that the atom graph can be computed from the set of atoms in $O(min (n^{\alpha},nm))$ time.
\section{Atom hypergraph} \label{sectAtomHypergraph}
%§§§§§§§§§§§§§§§§§§§§§§§§§§§§§§§§§§§§§§§§§§§§§§§§§§§§

In this section, we define the atom hypergraph of a graph and relate it to the more general notion of $\alpha$-acyclic hypergraph.

\begin{Definition}
Let $G = (V,E)$ be a graph. The {\em atom hypergraph} of $G$ is the hypergraph $H_A(G) = (V, {\cal A}(G))$.
\end{Definition}

Thus the atom trees of a connected graph are the join trees of its atom hypergraph.
We recall that for each hypergraph $H$, $2SEC(H)$ is the graph whose vertex set is the vertex set of $H$ and whose edges are the pairs of vertices that are contained in a hyperedge of $H$ 

\begin{Characterization} \label{carHAG}
An hypergraph is the atom hypergraph of a connected graph if and only if it is a connected $\alpha$-acyclic clutter, and in that case it is the atom hypergraph of the graph $2SEC(H)$ which is a connected chordal graph.
\end{Characterization}

\begin{Proof}
The atom hypergraph of a connected graph $G$ is connected (since $G$ is and each edge of $G$ is contained in an atom of $G$), $\alpha$-acyclic (since $G$ has an atom tree) and a clutter (by definition of atoms). Conversely, if $H$ is a connected $\alpha$-acyclic clutter then 
by Property~\ref{prop2SECchordal} it is the atom hypergraph  of the graph $2SEC(H)$ which is chordal, and which is connected since $H$ is.
\end{Proof}

Note that if $H$ is the atom hypergraph of $G$ then $2SEC(H)$ is the graph $G^+$ defined in Notation~\ref{notGplus}. Thus we refind that $G^+$ is chordal and has the same atoms as $G$ (Property~\ref{propG*Leimer}).

\begin{Definition}
The {\em union join graph} of an $\alpha$-acyclic hypergraph $H$, denoted by $UJ(H)$, is the union of its join trees.
\end{Definition}

\noindent
As the atom graph of a connected graph $G$ is the union of its atom trees by Characterizations~\ref{carAGunionAT}, we have the following property.

\begin{Property} \label{propAGUJ}
The atom graph of a connected graph is the union join graph of its atom hypergraph.
\end{Property}

 As a generalization Characterizations~\ref{carEdgeAG}, the union join graph of an $\alpha$-acyclic hypergraph $H$ can be computed from a join tree of $H$ by the following operation $tuj$, where $tuj$ stands for ``to union join''.

\begin{Definition}
For each join tree $T = ({\cal E}, E_T)$ of a hypergraph, $tuj(T)$ is the graph whose node set is ${\cal E}$ and whose edges are the pairs $\{X,Y\}$ of ${\cal E}$ such that there is an edge $X'Y'$ of $P_T(X,Y)$ such that $X \cap Y = X' \cap Y'$ (or equivalently $X' \cap Y' \subseteq X \cap Y$).
\end{Definition}

\begin{Characterization} \label{carUJtuj}
For each $\alpha$-acyclic hypergraph $H$ and each join tree $T$ of $H$, $UJ(H) = tuj(T)$.
\end{Characterization}

\begin{Proof}
Let $H = (V,{\cal E})$ and let $\{X,Y\} \subseteq {\cal E}$. Let us show that $XY$ is an edge of $UJ(H)$ if and only if $XY$ is an edge of $tuj(T)$. \\
$\Rightarrow$ : let $T'$ be a join tree of $H$ such that $XY$ is an edge of $T'$, and let ${\cal E}_X$ (resp. ${\cal E}_Y$) be the connected component of $T'-\{XY\}$ containing $X$ (resp. $Y$).
As $X \in {\cal E}_X$ and $Y \in {\cal E}_Y$, there is an edge $X'Y'$ of $P_T(X,Y)$ such that $X' \in {\cal E}_X$ and $Y' \in {\cal E}_Y$.
As $T'$ is a join tree and $XY$ is an edge of $P_{T'}(X',Y')$, $X' \cap Y' \subseteq X \cap Y$.
Hence $XY$ is an edge of $tuj(T)$. \\
$\Leftarrow$ : let $X'Y'$ be an edge of $P_T(X,Y)$ such that $X \cap Y = X' \cap Y'$, and let $T'$ be the graph $(T-\{X'Y'\})+ \{XY\}$. 
$T'$ is a tree having the same weight as $T$ (since $w(XY) = w(X'Y')$), so by Characterization~\ref{carMaxWeight} 
 $T'$ is also a join tree of $H$, and therefore $XY$ is an edge of $UJ(H)$.
\end{Proof}

Thus we refind Characterization~\ref{carEdgeAG} from Property~\ref{propAGUJ} and Characterization~\ref{carUJtuj}.
Conversely, Characterization~\ref{carUJtuj} can be deduced from  Characterization~\ref{carEdgeAG} and Property~\ref{propHtoAG} below, which shows that any $\alpha$-acyclic hypergraph is an atom hypergraph up to isomorphism.

\begin{Notation}
Let ${\cal E}$ and ${\cal E}'$ be two sets and let $f$ be a one-to-one mapping from ${\cal E}$ to ${\cal E}'$.
For each graph $K = ({\cal E},E_K)$, $f(K)$ denotes the graph obtained from $K$ by isomorphism $f$, i.e. $f(K) = ({\cal E}',\{f(X)f(Y), XY \in E_K\})$.
\end{Notation}

\begin{Property} \label{propHtoAG}
Let $H = (V,{\cal E})$ be an $\alpha$-acyclic hypergraph. Then there is a connected chordal graph $G = (V',E_G)$ and a one-to-one mapping $f$ from ${\cal E}$ to ${\cal A}(G)$ such that : \\
1) for each tree $T = ({\cal E},E_T)$, $T$ is a join tree of $H$ if and only if $f(T)$ is an atom tree of $G$, \\
2) $AG(G) = f(UJ(H)$, \\
3) If $H$ is connected then for each pair $\{X,Y\}$ of ${\cal E}$, $f(X) \cap f(Y) = X \cap Y$, otherwise there is an element $a$ of $V'$ such that for each pair $\{X,Y\}$ of ${\cal E}$, $f(X) \cap f(Y) = (X \cap Y)+\{a\}$, \\
4) for each join tree $T$ of $H$, $tuj(f(T)) =f(tuj(T))$.
\end{Property}

%ITEM 3 UTILE SEULEMENT POUR GRAPH-UNCOMP

\begin{Proof}
By Characterization~\ref{carHAG}, it is sufficient to find a connected $\alpha$-acyclic clutter $H' = (V',{\cal E}')$ 
and a one-to-one mapping $f$ from ${\cal E}$ to ${\cal E}'$ such that : 
(1) for each tree $T = ({\cal E},E_T)$, $T$ is a join tree of $H$ if and only if $f(T)$ is a join tree of $H'$, 
2) $UJ(H') = f(UJ(H)$,
and items 3) and 4).
Let ${\cal E}'$ be defined from ${\cal E}$ by adding a new specific element $a_X$ to each element of ${\cal E}$ which is not inclusion-maximal in ${\cal E}$, and adding a new common element $a$ to each element of ${\cal E}$ if $H$ is not connected.
Let $f$ map each element of ${\cal E}$ to the element of ${\cal E}'$ obtained from it, let $V' = \cup_{X \in {\cal E}'} X$, and let $H' = (V',{\cal E}')$. By definition, $H'$ is a connected  clutter satisfying 3).
As for each added element $a_X$ (resp. $a$) the set of elements of ${\cal E}'$ containing it is reduced to $\{X\}$ (resp. equal to ${\cal E}'$), $H'$ is $\alpha$-acyclic and satisfies (1). (2) follows from (1) and 4) follows from 3).
\end{Proof}
\par
Thus we can deduce from properties of $\alpha$-acyclic hypergraphs (proved from the definition of $\alpha$-acyclicity) properties of atom graphs, and conversely, we can deduce from properties of atom graphs (proved from properties of the minimal separators of the underlying graph) properties of general  $\alpha$-acyclic hypergraphs. This double approach helps to increase knowledge in both domains of atom graphs and  $\alpha$-acyclic hypergraphs, as some properties are easier to see in one of these domains than in the other one.
\par
We point out here the incoherence between the atom graph and the atom hypergraph of a non-connected graph, which comes from the choice of the definition of separators of a non-connected graph.
 For a non-connected $\alpha$-acyclic hypergraph $H$, a join tree of $H$ is defined from the join trees of the connected components of $H$ by adding edges (associated with the empty set) between these join trees to obtain a tree. We recall that according to the definition of separators given in this paper, we associate with each non-connected graph the forest of atom trees of its connected components. An alternative definition of separators, which is given for instance in \cite{Leimer}, would preserve the coherence between the graph and hypergraph approaches, 
 as well as Characterizations~\ref{carEdgeAtomTree} and \ref{carEdgeAG}. It defines a separator in the same way in a non-connected graph as in a connected one :
$S$ is an $ab$-separator of $G$ if $a$ and $b$ are in different connected components of $G(V \setminus S)$. 
It follows that the empty set is the unique minimal $ab$-separator of $G$ if $a$ and $b$ are in different connected components of $G$. Thus, according to this alternative definition, an atom tree of a (not necessarily connected) graph is a join tree of its atom hypergraph, and its atom graph is obtained from the atom graphs of its connected components by adding all edges between these atom graphs, as is the case for the union join graph of its atom hypergraph. 
Thus the results given in Section~\ref{sectComputeAG} for connected graphs extend to any graph when using this alternative definition of separators and to $\alpha$-acyclichypergraphs that are not necessarily connected, as will be seen in  Section~\ref{sectComputeUJ}.

\section{Computing the union join graph} \label{sectComputeUJ}

Algorithms and complexity results of Section~\ref{sectComputeAG} extend to the computation of the union join graph of an $\alpha$-acyclic hypergraph. They immediately extend to a connected $\alpha$-acyclic clutter $H$ since
 in that case $H$ is the atom hypergraph of $2SEC(H)$ by Characterization~\ref{carHAG}.
The algorithms  still hold for any $\alpha$-acyclic hypergraph since the proofs of their correctness do. It is also the case for the complexity bounds in function of parameters $n$, $p$, $s$ and $s_{\triangle}(T)$ whose definitions naturally extend to $\alpha$-acyclic hypergraphs as follows.

\begin{Notation}
For each $\alpha$-acyclic hypergraph $H = (V,{\cal E})$, 
$n = |V|$, $m$ is the number of edges of $2SEC(H)$, $\overline{m}$ is the number of edges of its complement, $p = |{\cal E}|$, $s = \Sigma_{X \in {\cal E}} |X|$, and for each join tree $T = ({\cal E},E_T)$ of $H$ $s_{\triangle}(T) = \Sigma_{XY \in {E_T}} |X \triangle Y|$.
\end{Notation}

%Note that $n \leq s$ (since $V = \cup_{X \in {\cal E}} X $),
%$p \leq s$ (since no hyperedge is the empty set), $p-1 \leq s_{\triangle}$ (since the hyperedges are pairwise distinct) and $s \leq np$.

We recall that $\alpha$ is the real number such that $O(n^{\alpha})$ is the best known time complexity of matrix multiplication and that the subset relation is defined in Definition~\ref{defsub}.

\begin{Theorem} \label{thComplexUJ}
The union join graph of an $\alpha$-acyclic hypergraph $H$ can be computed : \\
a) in $O(p^2)$ time from a join tree of $H$ and either its subset relation or the weighted line graph of $H$, \\
b) in $O(n + p^2)$ time from the weighted line graph of $H$, \\
c) in $O(min((n+p)^{\alpha},ps,p(n+s_{\triangle}(T))))$ time from a join tree $T$ of $H$, \\
d) in $O(min((n+p)^{\alpha},ps))$ time from $H$. \\
\end{Theorem}

\begin{Proof}
Item a)  follows from Theorem~\ref{thUJsub} and Theorem~\ref{thUJmin},
item b)  follows from Corollary~\ref{corUJmax},
item c) follows from item d) and the extension of Theorem~\ref{thAGdelta} to $\alpha$-acyclic hypergraphs,
and item d) follows from item b) and Proposition~\ref{propUJmax}.
\end{Proof}

The four results below extend Theorem~\ref{thAGsub}, Proposition~\ref{propAGsub}, Corollary~\ref{corAGmax} and Proposition~\ref{propAGmax} respectively. The complexity bound $n^{\alpha}$ is replaced by $(n+p)^{\alpha}$, which is the original bound appearing in the proofs of the concerned results and has been simplified into $n^{\alpha}$ since $p \leq n$ in the case of the atom graph (a graph has at most $n$ atoms).

\begin{Theorem} \label{thUJsub}
Given a join tree of an $\alpha$-acyclic hypergraph $H$ and its subset relation,
 Algorithm Forest Join computes the union join graph of $H$ in $O(p^2)$
 time.
\end{Theorem}

\begin{Proposition} \label{propUJsub}
Given a join tree of an $\alpha$-acyclic hypergraph, its subset relation can be computed in $O(min ((n+p)^{\alpha},ps))$ time.
\end{Proposition}

\begin{Corollary} \label{corUJmax}
(of Theorem~\ref{thUnionmax})
Given the weighted line graph of an $\alpha$-acyclic hypergraph $H$,
 Algorithm Union-max-weight computes the union join graph of $H$ in $O(n+p^2)$ time.
\end{Corollary}

\begin{Proposition} \label{propUJmax}
Given an $\alpha$-acyclic hypergraph, its weighted line graph can be computed in $O(min((n+p)^{\alpha},ps))$
 time.
\end{Proposition}

We present now Algorithm UJ-min-weight, which is an alternative to Algorithm Union-max-weight computing the union join graph in $O(p^2)$ time instead of $O(n+p^2)$ time, but requires a join tree of $H$ as input in addition to the weighted line graph of $H$. This algorithm obviously computes the atom graph of a connected graph, but Algorithm AG-max-weight already does it with the same complexity in $O(n^2)$ time and less input. The algorithm follows from Characterization~\ref{carmin} below, which is an immediate consequence of the characterization of $UJ(H)$ as $tuj(T)$ (Characterization~\ref{carUJtuj}). The algorithm computes for each pair $\{X,Y\}$ of hyperedges of $H$ the minimum weight of an edge of the path in $T$ between $X$and $Y$ and stores it in the variables $minWeight(X,Y)$ and $minWeight(Y,X)$ to be used later in the execution.

\begin{Characterization} \label{carmin}
Let $H = (V,{\cal E})$ be an $\alpha$-acyclic hypergraph, let $T$ be a join tree of $H$ and let $\{X,Y\} \subseteq {\cal E}$. Then $XY$ is an edge of $UJ(H)$ if and only if its weight is the minimum weight of an edge of $P_T(X,Y)$.
\end{Characterization}

\begin{Proof}
Let $w_{min}$ be the minimum weight of an edge of $P_T(X,Y)$. As $X \cap Y$ is a subset of $X' \cap Y'$ for each edge $X'Y'$ of $P_T(X,Y)$ since $T$ is a join tree, $w(XY) \leq w_{min}$, and $w(XY) = w_{min}$ if and only there is an edge $X'Y'$ of $P_T(X,Y)$ such that $X \cap Y = X' \cap Y'$, i.e. if and only if $XY$ is an edge of $UJ(H)$ by Characterization~\ref{carUJtuj}.
\end{Proof}

\begin{algorithm}[h!] 
\SetKwInOut{Input}{input}
\SetKwInOut{Output}{output}
\textbf{Algorithm UJ-min-weight}\vfill
\Input{A join tree $T = ({\cal E},E_T)$ and the weighted line graph $L_w(H)$ of an $\alpha$-acyclic hypergraph $H$.}
\Output{The union join graph of $H$.}
%\vspace{-0,1cm}
\BlankLine 
{ % begin 
// in the following, $w(e) = 0$ if $e$ is a non-edge of $L_w(H)$\;
Choose a node $X$ of $T$\;
$Reached \leftarrow \{X\}$; $Queue \leftarrow \{X\}$; $E' \leftarrow E_T$\;
\While{$Queue \neq \emptyset$}
{% while
Remove a node $X$ from $Queue$\;
\ForEach {$Y \in N_T(X)$}
{% foreach
\If {$Y \notin Reached$}
{% if
$minWeight(X,Y) \leftarrow w(XY)$; 
$minWeight(Y,X) \leftarrow w(XY)$\;
\ForEach {$Z \in Reached \setminus \{X\}$}
{% foreach2
$mw \leftarrow min(w(XY), minWeight(X,Z))$\;
$minWeight(Y,Z) \leftarrow mw$;
$minWeight(Z,Y) \leftarrow mw$\;
\If {$mw =  w(YZ)$}
{% if2
Add $YZ$ to $E'$\;
}% if2
}% foreach2
Add $Y$ to $Reached$ and to $Queue$\;
}% if
}% foreach
}% while
return $({\cal E},E')$\;
} % begin
\end{algorithm}

\begin{Theorem} \label{thUJmin}
Given a join tree and the weighted line graph of an $\alpha$-acyclic hypergraph $H$,
 Algorithm UJ-min-weight computes the union join tree of $H$ in $O(p^2)$ time.
\end{Theorem}

\begin{Proof}
Correctness follows from the fact that by Characterization~\ref{carmin} the following proposition $P$ is clearly an invariant of the main foreach and while loops. \\
$P$: $\forall \{X,Y\} \subseteq {\cal E}$, if $\{X,Y\} \subseteq Reached$ then
$(minWeight(X,Y)$ is the minimum weight of an edge of $P_T(X,Y) \wedge (XY \in E' \Leftrightarrow XY$ is an edge of $UJ(H)))$ otherwise $(XY \in E' \Leftrightarrow XY \in E_T)$. \\
The algorithm runs in $O(p^2)$ time, by numbering the elements of ${\cal E}$ from $1$ to $p$ and storing the values of $MinCard(X,Y)$  for each $(X,Y)$ in ${\cal E}^2$ such that $X \neq Y$ in an array $p \times p$.
\end{Proof}

By Characterization~\ref{carHAG} the complexity bounds in function of $n$,  $m$ and $\overline{m^+}$ presented in Section~\ref{sectComputeAG} extend to each connected $\alpha$-acyclic clutter $H$ 
replacing $\overline{m^+}$ by $\overline{m}$
 as the graph $G = 2SEC(H)$ is equal to $G^+$ since it is chordal (Property~\ref{prop2SECchordal}). 
In fact they also hold for each $\alpha$-acyclic clutter, replacing $m$ by $n+m$.
This follows from the fact that the bounds of the parameters $p$, $s$ and $s_{\triangle}(T)$ by functions of $n$,  $m$ and $\overline{m^+}$ extend to $\alpha$-acyclic clutters.

\begin{Property} \label{propclutter}
For each $\alpha$-acyclic clutter $H$, 
$p \leq n$, $s \leq n+m$, and for each join tree $T$ of $H$ $s_{\triangle}(T) \leq n + \overline{m}$.
\end{Property}

\begin{Proof}
By Characterization~\ref{carHAG} these inequalities hold if $H$ is connected.
It can be proved that they also hold if $H$ is disconnected by checking that the proofs of these inequalities given in Section~\ref{sectComputeAG} still hold. It can also be directly checked as follows. Let $H_1, \ldots , H_k$ the connected components of $H$, and for each $i$ in $[1,k]$ and each variable $v$ let $v_i$ be the value of $v$ in $H_i$. Then $p = \Sigma_{i=1}^k p_i \leq \Sigma_{i=1}^k n_i =n$.
Similarly $s \leq n+m$.
For $s_{\triangle}(T)$ we have $s_{\triangle}(T) = \Sigma_{i=1}^k s_{\triangle}(T_i) + nb_1$, where $nb_1 = \Sigma_{XY \in E_T, X \cap Y = \emptyset} |X|*|Y|$ and $\overline{m} = \Sigma_{i=1}^k \overline{m_i} + nb_2$, where $nb_2 = \Sigma_{\{i,j\} \subseteq [1,k] } |V_i|*|V_j|$. As $nb_1 \leq nb_2$, it follows that $s_{\triangle}(T) \leq n + \overline{m}$.
\end{Proof}

\begin{Corollary} \label{corclutter}
The complexity bounds in function of $n$,  $m$ and $\overline{m^+}$ presented in Section~\ref{sectComputeAG} hold for each $\alpha$-acyclic clutter $H$, replacing $m$ by $n+m$ and $\overline{m^+}$ by $\overline{m}$.
\end{Corollary}

If $H$ is an $\alpha$-acyclic hypergraph which is not a clutter, the values of $p$, $n$ and $s_{\triangle}(T)$ may be exponential in $n$. It is the case of the hypergraph $H = (V,P(V) \setminus \{\emptyset \})$,
which is $\alpha$-acyclic since $V$ is a hyperedge of $H$ (the tree whose edges are the pairs of hyperedges containing $V$ is a join tree of $H$).

%%%%%%%%%%%%%%%%%
\section{Conclusion}\label{conclusion}%Conclusion
%%%%%%%%%%%%%%%%%

In this paper, we provide two efficient algorithms to compute the atom graph of a graph, and extend them to compute the union join graph of an $\alpha$- acyclic hypergraph.
\par
Our algorithms, in the general case, compute the atom graph at no extra cost than computing the atoms.
\par
It would be interesting to explore the class of graphs which are isomorphic to atom graphs, and to provide a recognition algorithm for this class.

\textbf{Conflict of interest}

The authors declare no conflict of interest.

\end{document}